\documentclass[conference]{IEEEtran}
\usepackage{amsmath,graphicx,amsfonts,epsfig}
\usepackage{times}
\usepackage{amsbsy}
\usepackage{latexsym}
\usepackage{amssymb}
\begin{document}
\title{Distributed Space-Time Coding Techniques with Dynamic Buffers for Cooperative DS-CDMA Systems}


\author{\IEEEauthorblockN{Jiaqi Gu}
\IEEEauthorblockA{Communications Research Group\\Department of Electronics\\
University of York, U.K.\\
Email: jg849@york.ac.uk}
\and
\IEEEauthorblockN{Rodrigo C. de Lamare}
\IEEEauthorblockA{Communications Research Group\\Department of Electronics\\
University of York, U.K.\\CETUC, PUC--Rio, Brazil\\
Email: rcdl500@york.ac.uk}}

\maketitle

\begin{abstract}
In this work, we propose a dynamic buffer-aided distributed space-time
coding (DSTC) scheme for cooperative direct-sequence code-division
multiple access systems. We first devise a relay selection algorithm
that can automatically select the optimum set of relays among both
the source-relay phase and the relay-destination phase for DSTC
transmission according to the signal-to-interference-plus-noise
ratio (SINR) criterion. Multiple relays equipped with dynamic buffers are
introduced in the network, which allows the relays to store data
received from the sources and wait until the most appropriate time
for transmission. The proposed technique improves the
quality of the transmission with an acceptable delay as the buffer
size is adjustable. Simulation results show that the proposed
dynamic buffer-aided DSTC scheme and algorithm outperforms prior art.
\end{abstract}


\IEEEpeerreviewmaketitle

\section{Introduction}

The ever-increasing demand for performance and reliability in
wireless communication has encouraged the development of numerous
innovative techniques. Among them, cooperative diversity is one of
the key techniques that has been considered in recent years
\cite{Proakis} as an effective tool to improving transmission
performance and system reliability. Several cooperative schemes have
been proposed \cite{sendonaris,Venturino,laneman04,jpais,smce,tds},
and among the most effective ones are Amplify-and-Forward (AF),
Decode-and-Forward (DF) \cite{laneman04} and various distributed
space-time coding (DSTC) technique
\cite{Wornell,Yiu,RCDL3,Jing1,Peng,armo,bfdstc}. For an AF protocol,
relays cooperate and amplify the received signals with a given
transmit power amplifying their own noise. With the DF protocol,
relays decode the received signals and then forward the re-encoded
message to the destination. DSTC schemes exploit spatial and
temporal transmit diversity by using a set of distributed antennas.
With DSTC, multiple redundant copies of data are sent to the
receiver to improve the quality and reliability of data
transmission. Applying DSTC at the relays provides multiple
processed signal copies to compensate for the fading and
noise, helping to achieve the attainable diversity and coding gains and interference mitigation. 

In cooperative relaying systems, different strategies that employ multiple
relays have been recently introduced in \cite{Jing,Clarke,Talwar,Jiaqi1,Peng11,Jiaqi2}. The aim of
relay selection is to find the optimum relay or set of relays that results in the greatest improvement of reliability.
Recently, a new
cooperative scheme with buffers equipped at relays has been introduced and
analyzed in \cite{Zlatanov1,Zlatanov2,Krikidis,Ikhlef}. The main purpose is to
select the best link according to a given criterion. In \cite{Zlatanov1}, a
brief introduction of the buffer-aided relaying protocols for different
networks is carefully described and the some practical challenges are
discussed. Later, a further study of the throughput and diversity gain
of the buffer-aided system is introduced in \cite{Zlatanov2}. In
\cite{Krikidis}, a new selection technique that is able to achieve the full
diversity gain by selecting the strongest available link in every time slot is
detailed. In \cite{Ikhlef}, a max-max relay selection (MMRS) scheme for
half-duplex relays with buffers is proposed. In particular, relays with the
optimum source-relay links and relay-destination links are chosen and
controlled for transmission and reception, respectively.

In this work, we propose buffer-aided DSTC schemes and algorithms for cooperative
direct-sequence code-division multiple access (DS-CDMA) systems. In the proposed buffer-aided DSTC schemes, a relay pair selection
algorithm automatically selects the optimum set of relays according to the
signal-to-interference-plus noise ratio (SINR) criterion.   Specifically, the
proposed algorithms can be divided into two parts. Initially, a link combination
associated with the optimum relay group is selected, which determines if the
buffer is ready for transmission or reception. In the second part, DSTC is
performed from the selected relay combination to the destination when the
buffers are switched to the transmission mode. The direct transmission occurs between the source and the relay combination when the buffers are in
the reception mode. With dynamic buffers equipped at each of the relays, the proposed schemes take advantage of the high storage capacity, where multiple blocks of data can be stored so that the most appropriate ones can be selected at a suitable time instant. The key advantage of introducing the dynamic buffers in the system
is their ability to store multiple blocks of data according to a chosen criterion so that the most appropriate
ones can be selected at a suitable time instant with the highest efficiency.


This paper is organized as follows. In Section II, the system model is
presented. In Section III, the buffer-aided cooperative DSTC scheme is explained.
In Section IV, the dynamic buffer design is given and explained. In Section
V, simulation results are presented and discussed. Finally, conclusions are
drawn in Section VI.

\section{DSTC Cooperative DS-CDMA system model}
\begin{figure}[!htb]
\begin{center}
\def\epsfsize#1#2{0.8\columnwidth}
\epsfbox{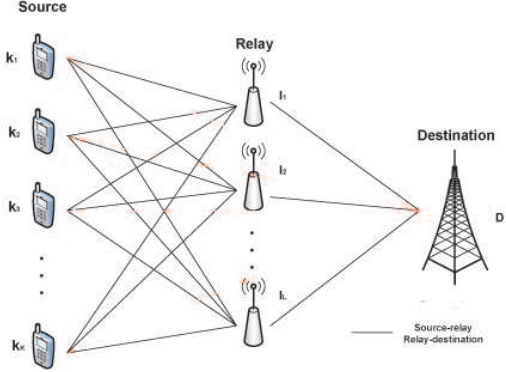} \vspace{-0.5em}\caption{\footnotesize
Uplink of a cooperative DS-CDMA system.} \vskip -5pt \label{fig1}
\end{center}
\vspace{-1em}
\end{figure}

We consider the uplink of a synchronous DS-CDMA system with $K$
users, $L$ relays equipped with finite-size buffers capable of storing $J$ packets and $N$ chips per
symbol that experiences flat fading channels. The system is equipped with a cooperative protocol at each relay and we assume that the
transmit data are organized in packets comprising $P$ symbols. The
received signals are filtered by a matched filter and sampled at chip
rate to obtain sufficient statistics.
The whole transmission is divided into two phases. In
the first phase, the source transmits the data to each of the relay over two consecutive time instants,
the decoded data over two time slots, $\hat{b}_{r_ld,k}(2i-1)$ and $\hat{b}_{r_ld,k}(2i)$, is stored at relay $l$
and is prepared to send data to the destination. A DSTC scheme is then employed at the following phase,
where the corresponding $2 \times 2$ Alamouti \cite{Alamouti,Chee,Liu} detected symbol matrix over relay $m$ and relay $n$ for user $k$ among two consecutive time instants is given by
\begin{equation}\label{equation0}
\textbf{B}_k = \left[\begin{array}{c c}
\hat{b}_{r_md,k}(2i-1) & -\hat{b}^*_{r_nd,k}(2i) \\
\hat{b}_{r_nd,k}(2i) & \hat{b}^*_{r_md,k}(2i-1) \\
 \end{array}\right].
\end{equation}
Consequently, the received signal from relay $m$ and $n$ to the destination
over two consecutive time slots yields the $2N\times1$ received vectors
described by
\begin{equation}\label{equation1}
\textbf{y}_{r_{m,n}d}(2i-1) = \sum\limits_{k=1}^K \textbf{h}_{r_md}^k
\hat{b}_{r_md,k}(2i-1)+ \textbf{h}_{r_nd}^k
\hat{b}_{r_nd,k}(2i)+\textbf{n}(2i-1),
\end{equation}
\begin{equation}\label{equation2}
\textbf{y}_{r_{m,n}d}(2i)=\sum\limits_{k=1}^K \textbf{h}_{r_nd}^k
\hat{b}^*_{r_md,k}(2i-1)- \textbf{h}_{r_md}^k
\hat{b}^*_{r_nd,k}(2i)+\textbf{n}(2i),
\end{equation}
where $\textbf{h}_{r_ld}^k=a_{r_ld}^k\textbf{s}_k h_{r_ld,k}$ denotes an $N \times 1$ effective signature vector for user $k$ from the $l$-th relay to the destination with $m,n
\in [1,2,...,L]$. The quantity $a_{r_ld}^k$ represents the $k$-th user's
amplitude from the $l$-th relay to the destination,
$\textbf{s}_k=[s_k(1),s_k(2),...s_k(N)]^T$ is the $N\times1$ signature sequence
for user $k$ and $h_{r_ld,k}$ are the complex channel fading coefficients for user $k$
from the $l$-th relay to the destination. The $N\times1$ noise vectors
$\textbf{n}(2i-1)$ and $\textbf{n}(2i)$ contain samples of zero mean complex
Gaussian noise with variance $\sigma^2$, $\hat{b}_{r_ld,k}(2i-1)$ and
$\hat{b}_{r_ld,k}(2i)$ are the decoded symbols at the output of relay $l$ after
using a cooperative protocol at time instants $(2i-1)$ and $(2i)$,
respectively. Equivalently, (\ref{equation1}) and (\ref{equation2}) can be
rewritten as
\begin{equation}
\textbf{y}_{r_{m,n}d}=\textbf{H}_{r_{m,n}d}^k \textbf{b}_{r_{m,n}d,k}+\textbf{n}_{r_{m,n}d},
\end{equation}
where $\textbf{y}_{r_{m,n}d}=\left[\textbf{y}^T_{r_{m,n}d}(2i-1),\textbf{y}^T_{r_{m,n}d}(2i)
\right]^T$ represents the received signal from relay $m$ and $n$ over two time
instants. The 2 N x 2 Alamouti matrix with the effective signatures for user k is given by
\begin{equation}
\textbf{H}_{r_{m,n}d}^k=\left[\hspace{-0.5em} \begin{array}{l}  \
\textbf{h}_{r_md}^k \ \ \ \ \ \ \ \ \  \textbf{h}_{r_nd}^k \\
\\
(\textbf{h}_{r_nd}^k)^* \ \ \ -(\textbf{h}_{r_md}^k)^* \\
\end{array}\hspace{-0.5em} \right],
\end{equation}
where $\textbf{h}_{r_ld}^k=a_{r_ld}^k\textbf{s}_k h_{r_ld,k}$
denotes the effective signature for user $k$ from the $l$-th relay
to the destination with $m,n\in [1,2,...,L]$. The $2\times1$ vector
$\textbf{b}_{r_{m,n}d,k}=\left[\hat{b}_{r_md,k}(2i-1),
\hat{b}_{r_nd,k}(2i)\right]^T$ is the processed vector when the DF
protocol is employed at relays $m$ and $n$ at the corresponding time
instant, and $\textbf{n}_{r_{m,n}d}=\left[  \textbf{n}(2i-1)^T,
\textbf{n}(2i)^T \right]^T$ is the noise vector that contains
samples of zero mean complex Gaussian noise with variance
$\sigma^2$. This scheme groups the relays into different pairs and a
more reliable transmission can be achieved if proper relay pair
selection is performed. Precoding techniques
\cite{cai2011switched,zu2012low,gsmi,mbthp,cai2015robust,rmbthp}
could also be used.

\section{proposed dynamic buffer-aided cooperative DSTC scheme}


In this section, we present a dynamic buffer-aided cooperative DSTC
scheme, where each relay is equipped with an adjustable buffer so
that the processed data can be stored and the buffer can wait until
the channel pair associated with the best performance is selected.
Consequently, processed data are stored at the corresponding buffer
entries and then re-encoded when the appropriate time interval
comes. Specifically, the size $J$ of the buffer is adjustable
according to a given criterion (e.g. the input SNR, channel
condition) so that a large amount of data can be eliminated from the
corresponding buffers and symbols can be sent directly or wait with
a shorter delay when the corresponding buffer size decreases. This
method effectively improves transmission reliability, and ensures
that the most suitable signal is selected from the buffer entries
and sent to the destination.

The algorithm begins with a SINR calculation for all possible channel
combinations. In the case of the Alamouti code, every two relays are combined
into a group and lists of all possible corresponding channel pairs are
considered. The SINR is then calculated and recorded as follows:
\begin{equation}\label{equation3}
\begin{split}
&{\rm SINR}_{sr_{m,n}}=\\
&\frac{\sum\limits_{k=1}^K \textbf{w}_{s_kr_m}^H \rho_{s_kr_m} \textbf{w}_{s_kr_m}+\textbf{w}_{s_kr_n}^H \rho_{s_kr_n} \textbf{w}_{s_kr_n} }{\sum\limits_{k=1}^K \sum\limits_{\substack{l=1\\l\neq m,n}}^L \textbf{w}_{s_kr_l}^H \rho_{s_kr_l} \textbf{w}_{s_kr_l} + \sigma^2\textbf{w}_{s_kr_m}^H \textbf{w}_{s_kr_m}+ \sigma^2\textbf{w}_{s_kr_n}^H \textbf{w}_{s_kr_n}},
\end{split}
\end{equation}
\begin{equation}\label{equation4}
\begin{split}
&{\rm SINR}_{r_{m,n}d}=\\
&\frac{\sum\limits_{k=1}^K (\textbf{w}_{r_md}^k)^H \rho_{r_md}^k \textbf{w}_{r_md}^k+(\textbf{w}_{r_nd}^k)^H \rho_{r_nd}^k \textbf{w}_{r_nd}^k}{\sum\limits_{k=1}^K \sum\limits_{\substack{l=1\\l\neq m,n}}^L (\textbf{w}_{r_ld}^k)^H \rho_{r_ld}^k \textbf{w}_{r_ld}^k + \sigma^2(\textbf{w}_{r_md}^k)^H \textbf{w}_{r_md}^k+ \sigma^2(\textbf{w}_{r_nd}^k)^H \textbf{w}_{r_nd}^k},
\end{split}
\end{equation}
where $\rho_{s_kr_l}=\textbf{h}_{s_kr_l}^H \textbf{h}_{s_kr_l}$ is the correlation coefficient of the desired user $k$ between the source and relay $l$, $\rho_{r_ld}^k=(\textbf{h}_{r_ld}^k)^H \textbf{h}_{r_ld}^k$ is the correlation coefficient for user $k$ from relay $l$ to the destination. $\textbf{h}_{s_kr_l}=a_{s_kr_l}\textbf{s}_kh_{s_kr_l}$ is the channel vector from user $k$ to relay $l$. In Eq. (\ref{equation3}), ${\rm SINR}_{sr_{m,n}}$ denotes the SINR for the combined paths from all
users to relay $m$ and relay $n$, $\textbf{w}_{s_kr_l}$ is the detector used at the relays. When the RAKE receiver is adopted at the corresponding relay, $\textbf{w}_{s_kr_l}$ is expressed as
\begin{equation}
\textbf{w}_{s_kr_l}=\textbf{h}_{s_kr_l},
\end{equation}
similarly, if the linear minimum mean-square error (MMSE) receiver \cite{RCDL5} is employed at the relays, $\textbf{w}_{s_kr_l}$ is equal to
\begin{equation}
\textbf{w}_{s_kr_l}=\bigg(\sum\limits_{k=1}^K \textbf{h}_{s_kr_l}\textbf{h}^H_{s_kr_l}+\sigma^2 \textbf{I}\bigg)^{-1}\textbf{h}_{s_kr_l},
\end{equation}
similarly, in Eq. (\ref{equation4}), ${\rm SINR}_{r_{m,n}d}$ represents the
SINR for the combined paths from relay $m$ and relay $n$ to the destination. The receive filter $\textbf{w}_{r_ld}^k$ is employed by the detector used at the destination. When the RAKE receiver is adopted at the destination, $\textbf{w}_{r_ld}^k$ is expressed as
\begin{equation}
\textbf{w}_{r_ld}^k=\textbf{h}_{r_ld}^k.
\end{equation}
Similarly, if the linear MMSE receiver
\cite{int,Chen,Meng,l1cg,zhaocheng,alt,jiolms,jiols,jiomimo,jidf,fa10,smtvb,saabf,barc,honig,mswfccm,song,locsme}.
is employed at the relays, $\textbf{w}_{r_ld}^k$ is equal to
\begin{equation}
\textbf{w}_{r_ld}^k=\bigg(\sum\limits_{k=1}^K \textbf{h}_{r_ld}^k(\textbf{h}_{r_ld}^k)^{H}+\sigma^2 \textbf{I}\bigg)^{-1} \textbf{h}_{r_ld}^k.
\end{equation}
The above equations correspond to a cooperative system  under the
assumption that signals from all users are transmitted to the
selected relays $m$ and $n$. Both RAKE and MMSE receivers are
considered here for the purpose of complexity, it should be
mentioned that other detectors
\cite{zhang2015large,de2013massive,mber,stbcccm,itic,RCDL4,jiomimo,spa,mbsic,mfsic,dfcc,mbdf,did,uchoa2016iterative}
can also be used. We then sort all these SINR values in a decreasing
order and select the one with the highest SINR as given by
\begin{equation}
{\rm SINR_{p,q}={\rm arg}\max _{\substack{m,n\in[1,2,...,L]}} \{ {\rm SINR}_{sr_{m,n}},{\rm SINR}_{r_{m,n}d} \}}, \label{equation5}
\end{equation}
where ${\rm SINR_{p,q}}$ denotes the highest SINR associated with the relay $p$ and relay $q$.
After the highest SINR corresponding to the combined paths is selected,
two different situations need to be considered as follows.\\
\\
\textbf{Source-relay link}:
\vspace{3mm}

If the highest SINR belongs to the source-relay link, then the signal sent to
the target relays $p$ and $q$ over two time instants is given by
\begin{equation}\label{equation6}
\textbf{y}_{sr_l}(2i-1)=\sum\limits_{k=1}^K \textbf{h}_{s_kr_l}b_k(2i-1)+\textbf{n}(2i-1), l\in[p,q],
\end{equation}
\begin{equation}\label{equation7}
\textbf{y}_{sr_l}(2i)=\sum\limits_{k=1}^K \textbf{h}_{s_kr_l}b_k(2i)+\textbf{n}(2i),l\in[p,q].
\end{equation}
The received signal is then processed by the detectors as the DF protocol is
adopted. Therefore, the decoded symbols that are stored and sent to the
destination from the $l$-th relay are obtained as
\begin{equation}
\hat{b}_{r_ld,k}(2i-1)=Q(\textbf{w}_{s_kr_l}^{H}\textbf{y}_{sr_l}(2i-1)),
\end{equation}
and
\begin{equation}
\hat{b}_{r_ld,k}(2i)=Q(\textbf{w}_{s_kr_l}^{H}\textbf{y}_{sr_l}(2i)),
\end{equation}
where $Q(\cdot)$ denotes the slicer. After that, the buffers
are switched to the reception mode, the decoded symbol is consequently
stored in the corresponding buffer entries. Clearly, these operations are
performed when the corresponding buffer entries are not full, otherwise, the
second highest SINR is chosen as given by
\begin{equation}\label{equation8}
{\rm SINR^{pre}_{p,q}}={\rm SINR_{p,q}}
\end{equation}
\begin{equation}\label{equation9}
{\rm SINR_{u,v}} \in {\rm max} \{ {\rm SINR_{sr_{m,n}}},  {\rm SINR_{r_{m,n}d}}\} \setminus {\rm SINR^{pre}_{p,q}},
\end{equation}
where ${\rm SINR_{u,v}}$ denotes the second highest SINR associated with the
updated relay pair $\Omega_{u,v}$. $\{ {\rm SINR_{sr_{m,n}}},  {\rm
SINR_{r_{m,n}d}}\} \setminus {\rm SINR^{pre}_{p,q}}$ denotes a complementary
set where we drop the ${\rm SINR^{pre}_{p,q}}$ from the link SINR set $\{ {\rm
SINR_{sr_{m,n}}},  {\rm SINR_{r_{m,n}d}}\}$. Consequently, the above process
repeats in the following time instants.
\\
\\
\textbf{Relay-destination link}:
\vspace{3mm}

If the highest SINR is selected from the relay-destination link, in the
following two consecutive time instants, the buffers are switched to
transmission mode and the decoded symbol for user $k$ is re-encoded with the
Alamouti matrix as in (\ref{equation0}) so that DSTC is performed from the
selected relays $p$ and $q$ to the destination as given by
\begin{equation}\label{equation10}
\textbf{y}_{r_{p,q}d}(2i-1)=\sum\limits_{k=1}^K \textbf{h}_{r_pd}^k \hat{b}_{r_pd,k}(2i-1)+
\textbf{h}_{r_qd}^k \hat{b}_{r_qd,k}(2i)+\textbf{n}(2i-1),
\end{equation}
\begin{equation}\label{equation11}
\textbf{y}_{r_{p,q}d}(2i)=\sum\limits_{k=1}^K \textbf{h}_{r_qd}^k \hat{b}^*_{r_pd,k}(2i-1)-
\textbf{h}_{r_pd}^k \hat{b}^*_{r_qd,k}(2i)+\textbf{n}(2i).
\end{equation}
The received signal is then processed by the detectors at the destination.
Clearly, the above operation is conducted under the condition that the
corresponding buffer entries are not empty, otherwise, the second highest SINR
is chosen according to (\ref{equation8}) and (\ref{equation9}) and the above
process is repeated.

\section{Proposed dynamic buffer scheme}
The size $J$ of the buffers plays a key role in the performance
of the system, which improves with the increase of the size as buffers with
greater size allow more data packets to be stored. In this case, extra degrees
of freedom in the system or choices for data transmission are available. Hence, in this section, we release the limitation on the size of the buffer to further explore the additional advantage brought by dynamic buffer design where the buffer size can vary according to different criteria such as the input SNR and the channel condition.
When considering the input SNR, extra buffer space is required when the transmission is performed in the low SNR region so that the data associated with the best channels can be selected among a greater number of candidates. On the other hand, in the high SNR region, a small buffer size is employed as most of the processed symbols are appropriate when compared with the situation in the low SNR region. In this work, we assume that the buffer size $J$ is inversely proportional to the input SNR, namely, with the increase of the SNR, the buffer size decreases automatically. The algorithm for calculating the buffer size $J$ is detailed in Table. \ref{table3}.
\begin{table}[!htb]
\centering\caption{The algorithm to calculate the buffer size $J$}
\begin{tabular}{l}
\hline
(1) \textbf{If}\ \ \ \ \ \ \ ${\rm SNR_{cur}}={\rm SNR_{pre}}+d_1$ \\
\\
(2) \textbf{then} \ \ \ \  $ J_{\rm cur}=J_{\rm pre}-d_2$,\\
\\
where ${\rm SNR_{cur}}$ and ${\rm SNR_{pre}}$ represent the input SNR after and before\\
increasing its value,\\
\\
$J_{\rm cur}$ and $J_{\rm pre}$ denote the corresponding buffer size before and after\\
decreasing its value,\\
\\
$d_1$ and $d_2$ are the step sizes for the SNR and the buffer size, respectively.\\
\hline
\end{tabular}\label{table3}
\end{table}\\

The buffer size can be determined by the current selected channel pair condition. In particular, we set a threshold $\gamma$ that denotes the channel power, if the current selected channel power is under $\gamma$, the buffer size increases as more candidates need to be saved in order to select the best symbol, on the contrary, if the current selected channel pair power exceeds $\gamma$, we decrease the buffer size as there is a high possibility that the transmission is not significantly affected. The approach based on the channel power for varying the buffer can be summarized in Table. \ref{table4}.
\begin{table}[!htb]
\centering\caption{The algorithm for calculate buffer size $J$ based on the channel power}
\begin{tabular}{l}
\hline
(1) \textbf{If}\ \ \ \ \ \ $\min\| h_{s_kr_l}\|^2 \leq \gamma$ \ \ \  \textbf{or} \ \ \ $\min\| h_{r_ld}\|^2 \leq \gamma, \ \ l\in[1,L]$ \\
\ \ \ \ \ \ \ \ \ $J_{\rm cur}=J_{\rm pre}+d_3$ \\
(2) \textbf{else}\\
\ \ \ \ \ \ \ \ \ $J_{\rm cur}=J_{\rm pre}-d_3$\\
\textbf{end}\\
\\
where $d_3$ represents the step size when adjusting the buffer size.\\
\hline
\end{tabular}\label{table4}
\end{table}\\

Then, we analyse the computational complexity required by the proposed relay pair selection algorithm. The exhaustive relay
pair search requires $(7KNL^3-7KNL^2)$ multiplications and
$(2KNL^3-2KNL^2+KL^3-KL^2-2L^2+2L)$ additions,
while the proposed greedy relay pair selection algorithm only
requires less than $(21KNL^2-7KNL)$ multiplications and
$(6KNL^2+3KL^2-3KL-L+1)$ additions, which is
an order of magnitude less costly. Therefore, when a large
number of relays participate in the transmission, with a careful
control of the buffer size $J$, a good balance of complexity and
performance is achieved.

At last, we analyze the average delay of the proposed schemes and algorithms. The major delay comes from three aspects. Firstly, the improvement of the performance brought by the buffer--aided
relays comes at the expense of the transmission delay \cite{Islam}. Secondly, the DSTC scheme will introduce further delay as the DSTC scheme takes two time--slots to transmit two packets in a time \cite{Bouanen,Gong,Sheng}. Finally, extra computational delay is required as the proposed relay selection algorithms are conducted in the transmission.

\section{simulations}
In this section, a simulation study of the proposed
buffer-aided DSTC technique for cooperative systems is carried out.
The DS-CDMA network uses randomly generated spreading codes of
length $N=16$. The corresponding channel coefficients are taken as
uniformly random variables and are normalized to ensure the total
power is unity for all analyzed techniques. We assume perfectly
known channels at the receivers. Equal power allocation with
normalization is assumed to ensure no extra power is introduced
during the transmission. We consider packets with 1000 BPSK symbols and step size $d=2$ when conducting the dynamic schemes. We consider fixed buffer--aided exhaustive (FBAE)/fixed buffer--aided greedy (FBAG) \cite{Jiaqi3} relay pair selection strategies (RPS) and dynamic buffer--aided exhaustive (DBAE)/dynamic buffer--aided greedy(DBAG) \cite{Jiaqi3} RPS, and we assumed perfect channel state information is available at the relays and the destination and that the performance of the system with channel estimation algorithms is slightly degraded.

\begin{figure}[!htb]
\centerline{
\includegraphics[width=0.5\columnwidth,height=0.85\columnwidth]{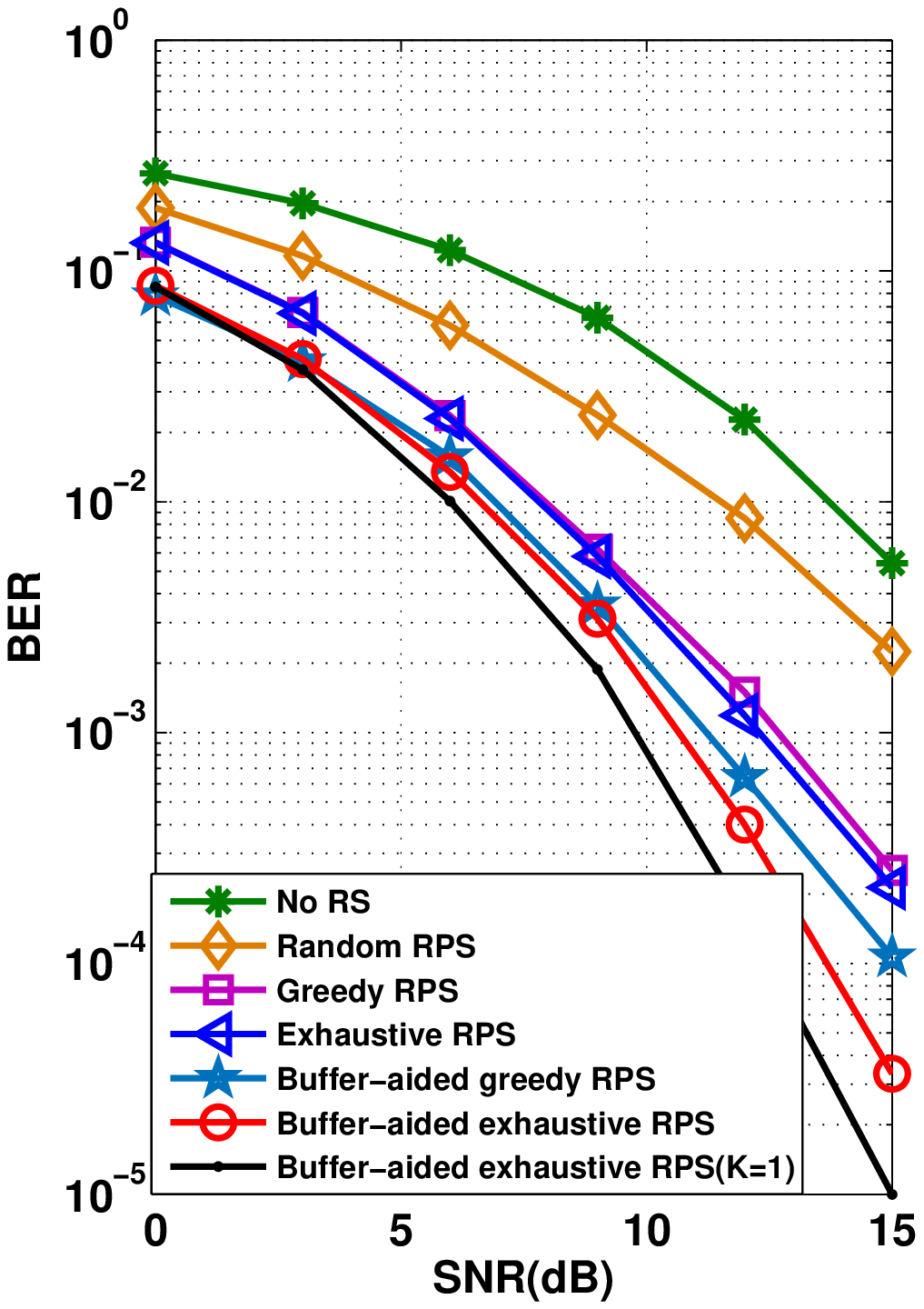}
\includegraphics[width=0.5\columnwidth,height=0.85\columnwidth]{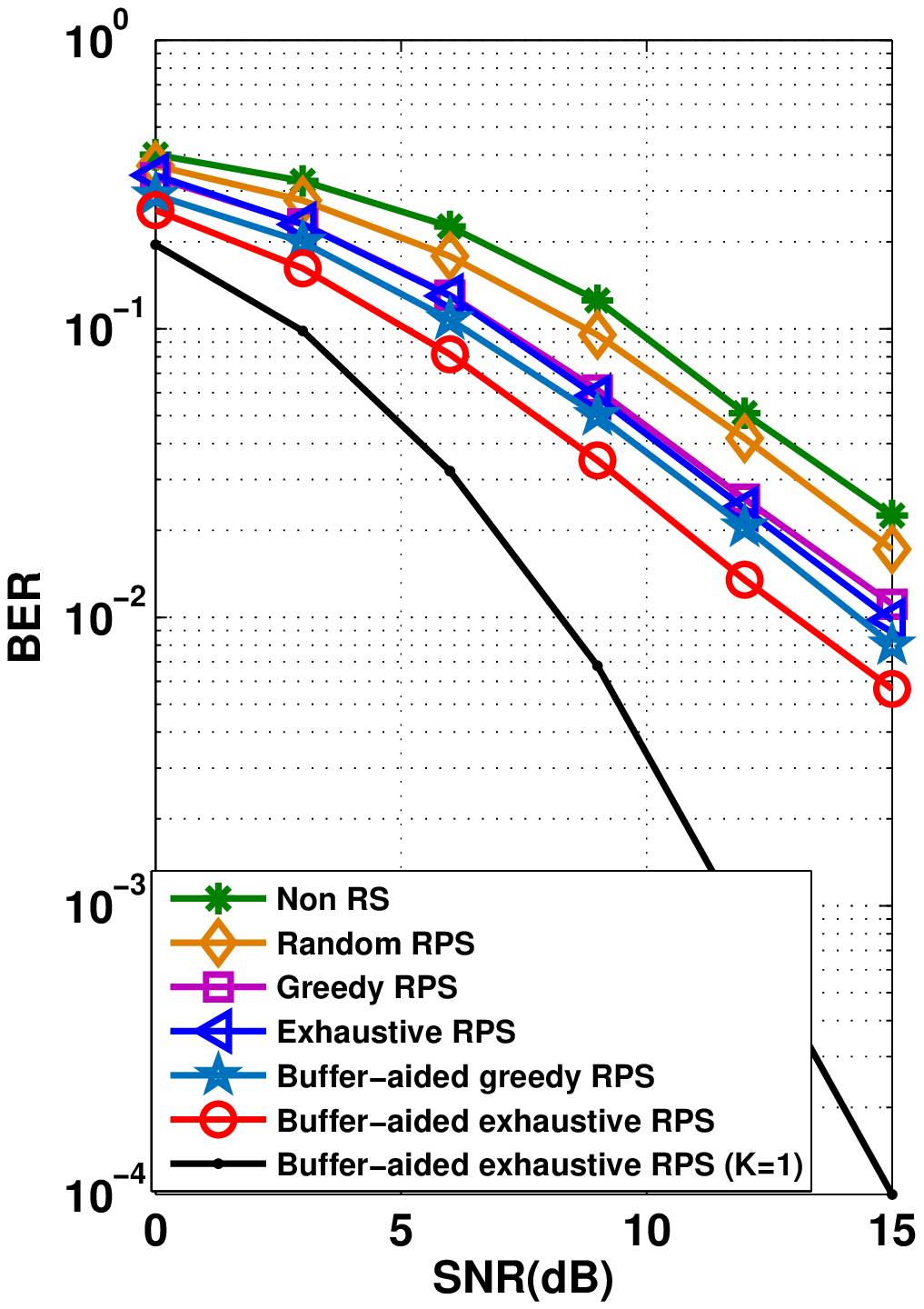}
} \vspace{-0.75em}\caption{\footnotesize a)\ Performance comparison
for buffer-aided scheme and non buffer-aided scheme in cooperative
DS-CDMA system with perfect decoding at the relay, RAKE at the
destination\ \ \ b)\ Performance comparison for buffer-aided scheme
and non buffer-aided scheme in cooperative DS-CDMA system with MMSE
at the relay, RAKE at the destination} \vspace{-0.2em} \label{fig2}
\end{figure}

In order to verify that the fixed buffer-aided relay pair DSTC cooperative scheme contributes to the performance gain, we compare the performance between the situations of the transmission with fixed size buffers and without buffers in Fig. \ref{fig2}. The first example shown in Fig. \ref{fig2}(a) illustrates the performance comparison between the proposed buffer-aided DSTC transmission with different RPS and DSTC transmission with relay pair selections and
no buffers when better decoding techniques are adopted. The system has 3 users,
6 relays, perfect decoding is assumed  at each relay and the matched filter is
adopted at the destination. Specifically, for the no relay selection (RS) DSTC
technique, all relays participate in the DSTC transmission (every two
consecutive relays are working in pairs). Similarly, for the non buffer-aided
schemes, the RPS process only occurs during the second phase
(relay-destination), where the random selection algorithm chooses an arbitrary
relay pair, the proposed greedy algorithm chooses two relays associated with
two optimum relay-destination links and the exhaustive relay pair schemes
examines all possible relay pairs and selects the one with the highest SINR. In
contrast, the proposed buffer-aided scheme automatically selects the relay pair
over both source-relay links and relay-destination links. Moreover, with the
help of the buffers, the most appropriate data are sent and better overall
system performance can be achieved. The performance for a single-user
buffer-aided exhaustive RPS DSTC is presented here for comparison purposes.
Consequently, the results reveal that our proposed buffer-aided strategies
($J=6$) perform better than the one without buffers. In particular, Fig. \ref{fig2}(a)
also illustrates that our proposed buffer-aided schemes can approach the single
user bound very closely.

The second example depicted in Fig. \ref{fig2}(b) compares the proposed
buffer-aided DSTC transmission with different RPS
and DSTC transmission with relay pair selections and no buffers. In
this scenario, where we apply the linear MMSE receiver at each of the relay and the RAKE at the
destination in an uplink cooperative scenario with 3 users, 6 relays
and buffer size $J=6$. Similarly, the performance bounds for a
single-user buffer-aided exhaustive RPS DSTC are presented for
comparison purposes. The results also indicate that our proposed
buffer-aided strategies ($J=6$) perform better than the one without
buffers. Furthermore, the BER performance curves of our greedy RPS
algorithm approaches the exhaustive RPS, while keeping the complexity
reasonably low for practical use.

\begin{figure}[!htb]
\centerline{
\includegraphics[width=0.5\columnwidth,height=0.8\columnwidth]{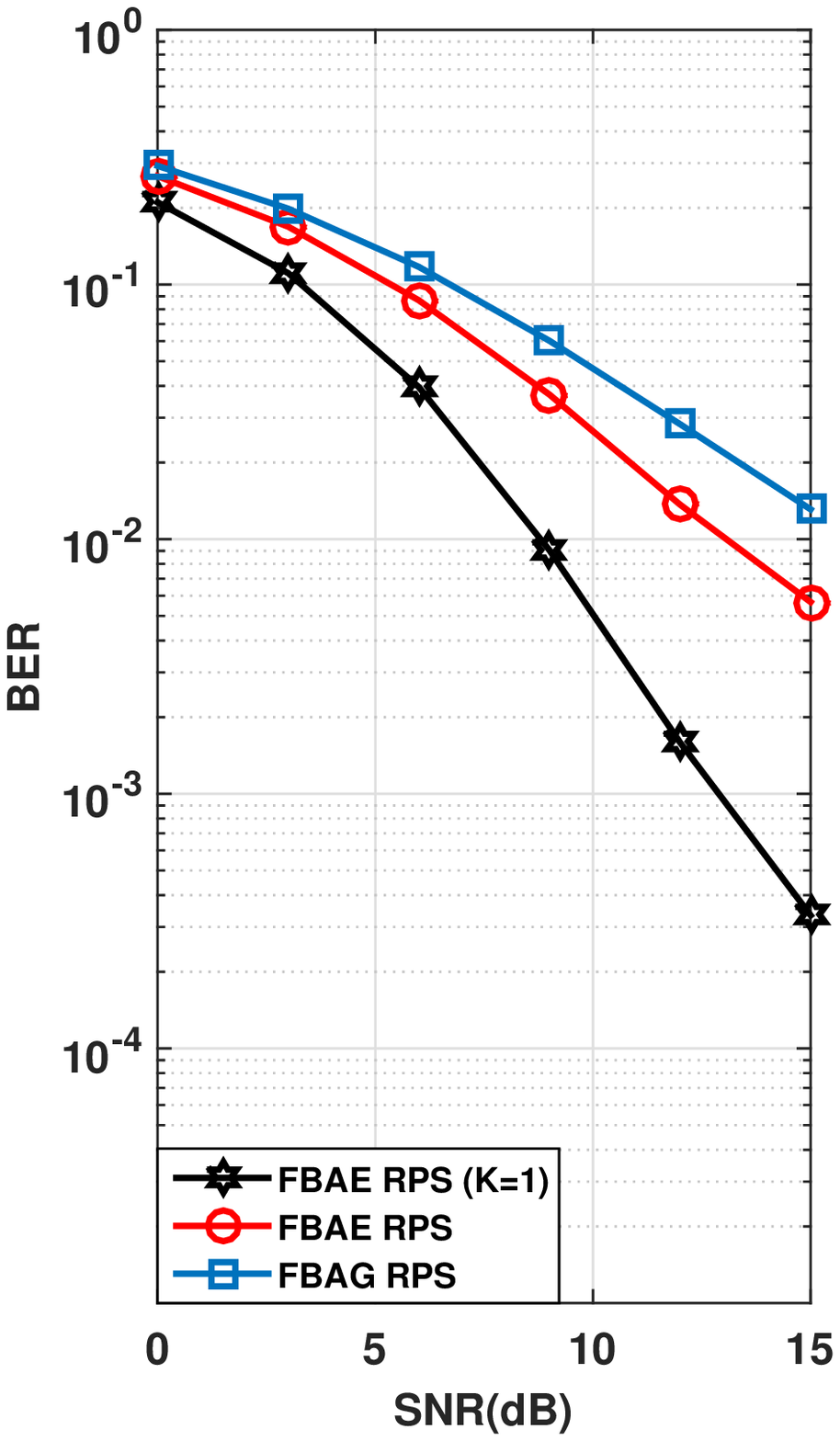}
\includegraphics[width=0.5\columnwidth,height=0.8\columnwidth]{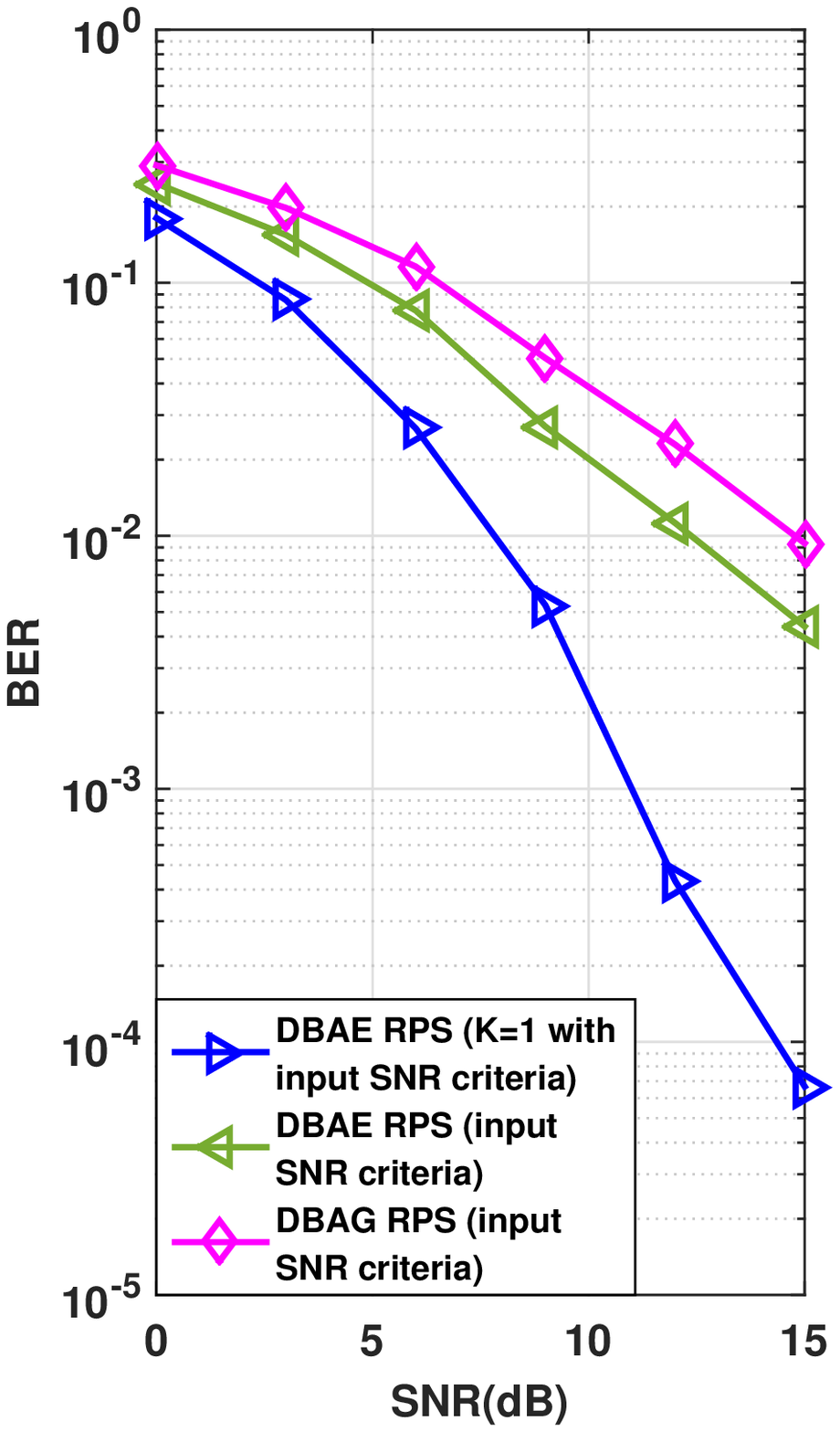}
} \vspace{-0.75em}\caption{\footnotesize a) performance comparison for fixed buffer design (input SNR criterion)\ \ \
b) performance comparison for dynamic buffer design (input SNR criterion)} \vspace{-0.2em} \label{fig3}
\end{figure}

The second example illustrates the performance comparison for the fixed buffer-aided design in Fig. \ref{fig3}(a) and dynamic buffer-aided design in Fig. \ref{fig3}(b) in a cooperative DSTC system with different relay pair selection strategies (RPS). The overall network has 3 users, 6 relays, the linear MMSE receiver is applied at each relay and the matched filter is adopted at the destination. For dynamic algorithms, the buffer size $J$ decreases when approaching higher SNR region. In both figures, the buffer-aided exhaustive greedy RPS algorithm performs better than the greedy one.

When we compare the curves in Fig. \ref{fig3}, the dynamic buffer techniques are more flexible than the fixed buffer ones as they explore the most suitable buffer size for the current transmission according to a given criterion. In this case, there is a greater possibility to select the most appropriate data when the transmission is operated in poor condition as more candidates are stored in the buffer space. On the other hand, the transmission delay can be avoided when the outer condition improves as most of the candidates are appropriate. Simulation results verify these points and indicate that the DBAE/DBAG RPS outperform the FBAE/FBAG ($J=8$) RPS and the advantage increases when adopting the single user case. Furthermore, it can also be seen that the BER performance curves of the greedy relay pair selection algorithm \cite{jiaqi} approaches the exhaustive search, whilst keeping the complexity reasonably low for practical utilization.

\begin{figure}[!htb]
\centerline{
\includegraphics[width=0.5\columnwidth,height=0.8\columnwidth]{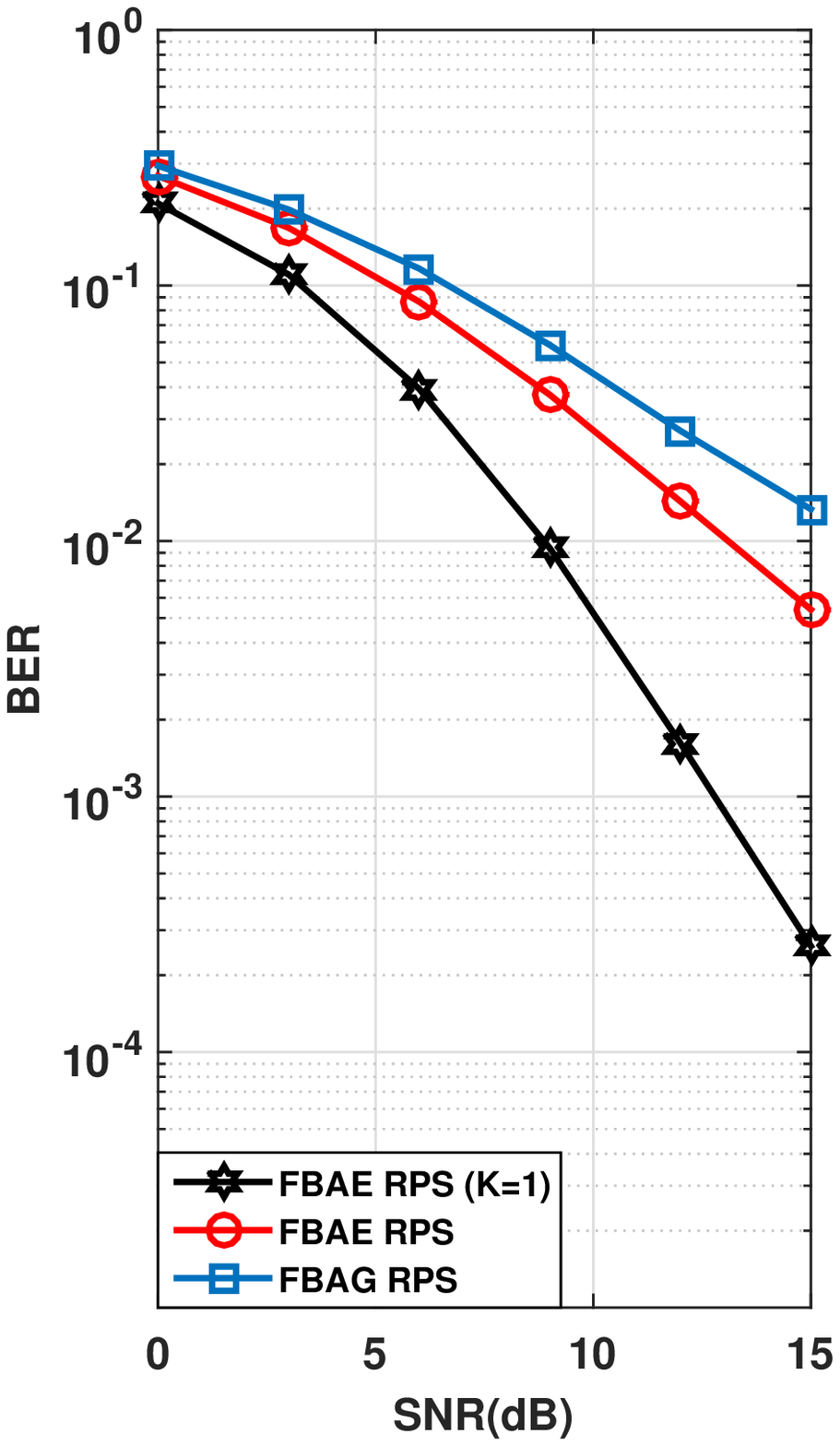}
\includegraphics[width=0.5\columnwidth,height=0.8\columnwidth]{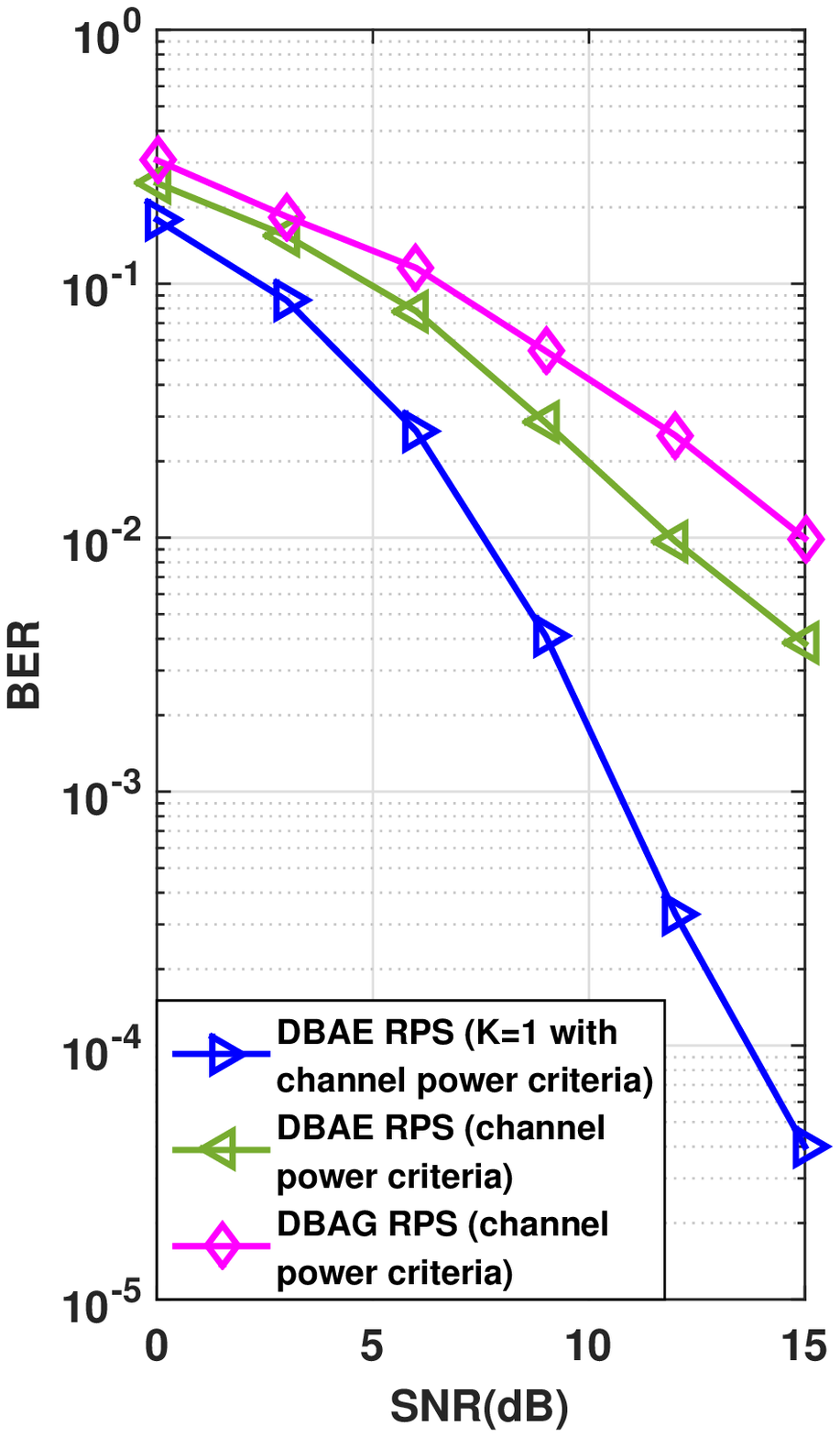}
} \vspace{-0.75em}\caption{\footnotesize a) performance comparison for fixed buffer design (channel power criterion)\ \ \ b)performance comparison for dynamic buffer design (channel power criterion)} \vspace{-0.2em} \label{fig4}
\end{figure}

The third example compares the FBAE/FBAG RPS scheme in Fig. \ref{fig4}(a) and the DBAE/DBAG RPS strategy in Fig. \ref{fig4}(b) in a DSTC cooperative system, where we apply the linear MMSE receiver at each of the relay and the matched filter (MF or RAKE receiver) at the destination in an uplink cooperative scenario with 3 users, 6 relays and fixed buffer size $J=8$. Similarly, the performance for a single-user buffer-aided exhaustive RPS DSTC is presented for comparison purposes. In both figures, the buffer-aided exhaustive search RPS algorithm performs better than the greedy one. The average dynamic buffer size $J$ is highly dependant on the threshold $\gamma$ and the step size $d$, clearly, with careful control on these parameters, better performance can be achieved. The simulation results also indicate that our proposed dynamic design perform better than the fixed buffer size ones when we apply the same relay selection method, as depicted in Fig. \ref{fig4}.


\section{conclusions}
In this work, we have presented a dynamic buffer-aided DSTC
scheme for cooperative DS-CDMA systems with different relay pair
selection techniques. With the help of the dynamic buffers, this approach
effectively improves the transmission performance and help to achieve a good balance between bit error rate (BER) and delay.
Simulation results show that
the performance of the proposed dynamic design can offer good gains
as compared to fixed buffer-aided schemes.

{\footnotesize
\bibliographystyle{IEEEbib}
\bibliography{reference}}

\end{document}